\documentclass[aps,twocolumn,showpacs]{revtex4}
\usepackage{bm}
\usepackage{color} 
\usepackage{graphicx}
\usepackage{epsfig}

\newcommand{\Ok}{\boldsymbol{\Omega}}
\newcommand{\av}[1]{\left< {#1} \right>}

\begin{document}

\title{Spin-orbit Hanle effect in high-mobility quantum wells}

\author{A.\,V.\,Poshakinskiy and S.\,A.\,Tarasenko}

\affiliation{A.F.~Ioffe Physical-Technical Institute, Russian Academy of Sciences, 194021 St.~Petersburg, Russia}


%72.25.Rb Spin relaxation and scattering  
%72.25.Fe Optical creation of spin polarized carriers  
%78.67.De Optical properties of Quantum wells  
%73.21.Fg Electron states and collective excitations in Quantum wells 
%73.20.-r Surface and interface electron states, Electronic structure of bulk materials 
%71.70.Ej Spin-orbit coupling, Zeeman and Stark splitting, Jahn-Teller effect  

\begin{abstract}
We study the depolarization of optically oriented electrons in quantum wells subjected to an in-plane magnetic field and show that the Hanle curve drastically depends on the carrier mobility. In low-mobility structures, the Hanle curve is described by a Lorentzian with the width determined by the effective $g$-factor and the spin lifetime. In contrast, 
the magnetic field dependence of spin polarization in high-mobility quantum wells is nonmonotonic: The spin polarization rises with the magnetic field induction at small fields, reaches maximum and then decreases. We show that the position of the Hanle curve maximum can be used to directly measure the spin-orbit Rashba/Dresselhaus magnetic field.
\end{abstract}

\maketitle

%%%%%%%%%%%%%%%%%%%%%%%%%%%%%%%%%%%%%%%%%%%%%%%%%%%%%%%%%%%%%%%%%%%%%%%%%%%
%\section{Introduction}
%%%%%%%%%%%%%%%%%%%%%%%%%%%%%%%%%%%%%%%%%%%%%%%%%%%%%%%%%%%%%%%%%%%%%%%%%%%

The absorption of circularly polarized light in semiconductor structures is known to result in spin orientation of photocarriers, which leads, in turn, to the circular polarization of recombination radiation~\cite{OO_book, Kusraev_ed,Dyakonov_ed}. The polarization of secondary radiation can be suppressed by applying the magnetic field perpendicularly to the initial light beam. Such an effect, first described by W. Hanle in 1920s for quicksilver atoms~\cite{Hanle24}, enables physicists to obtain valuable information on the spin lifetime and effective $g$-factor. Typically, the Hanle curve is described by the Lorentz function that reaches maximum at zero magnetic field and monotonically decays with the field increase. It was also shown that
the dependence of spin polarization on the external magnetic field can be anomalous if electron or hole spins are affected by additional magnetic moments of nuclei~\cite{OO_book,Krebs10} or magnetic ions~\cite{Kudinov03}. The Hanle curves measured in such systems do not follow the Lorentz function and can be even nonmonotonic. 

In this Letter, we study the effect of an in-plane magnetic field on spin dynamics of optically oriented electrons in quantum wells (QWs). We show that the form of Hanle curve drastically depends on the electron gas mobility even in non-magnetic structures. In low-mobility quantum wells, the Hanle curve is described by a Lorentzian with the width determined by the effective electron $g$-factor and spin lifetime. Surprisingly, the magnetic field dependence of spin polarization in high-mobility structures is nonmonotonic: The polarization rises with the field induction at small fields, reaches maximum and then decreases. Such a behavior is caused by spin-orbit splitting of electron states. The position of the polarization maximum can be used to directly measure the effective Rashba or Dresselhaus magnetic field.

%%%%%%%%%%%%%%%%%%%%%%%%%%%%%%%%%%%%%%%%%%%%%%%%%%%%%%%%%%%%%%%%%%%%%%%%%%%
%\section{Microscopic theory}
%%%%%%%%%%%%%%%%%%%%%%%%%%%%%%%%%%%%%%%%%%%%%%%%%%%%%%%%%%%%%%%%%%%%%%%%%%%

We consider a common $n$-type QW structure grown along the crystallographic axis $z \parallel [001]$
which is subjected to an in-plane magnetic field $\bf B$. The structure is excited by normally-incident circularly polarized light which creates spin polarized electrons at the rate $\bf G$.  The geometry is sketched in the inset of Fig.~1(b). 
In $(001)$ QWs, the spin relaxation of electrons is determined, in the wide range of temperature and carrier density, by the D'yakonov-Perel' spin relaxation mechanism~\cite{DP71,DK86}. The mechanism is based on the precession of individual electron spins in the effective Dresselhaus and/or Rashba magnetic fields originating from spin-orbit interaction in non-centrosymmetric structures (for a review, see Refs.~\cite{Averkiev08,Wu10}). In the framework of
kinetic theory, the steady-state distribution function $\bf s_k$ of electron spins in $\bf k$-space is described by equation
\begin{equation}\label{kinetic}
{\bf s}_{{\bf k}} \times ( \boldsymbol{\Omega}_L + \boldsymbol{\Omega}_{{\bf k}} ) = {\bf g}_{{\bf k}} - \frac{{\bf s}_{{\bf k}} - \langle {\bf s}_{{\bf k}} \rangle}{\tau} \:. 
\end{equation}
Here, $\boldsymbol{\Omega}_L$ is the Larmor frequency corresponding to the external field ${\bf B}$, its components are given by $\Omega_{L,\alpha} = (\mu_B/\hbar) \sum_\beta g_{\alpha\beta} B_{\beta} $, $g_{\alpha\beta}$ is the $g$-factor tensor, $\bm{\Omega_k}$ is the precession frequency in
the effective magnetic field, ${\bf g}_{{\bf k}}$ is the spin generation rate in the states with the wave vector ${\bf k}$,
$\tau$ is the isotropization time of the spin distribution function, and the angular brackets denote averaging over the direction of ${\bf k}$. We assume that resident electrons form the degenerate two-dimensional gas and ${\bf g}_{{\bf k}}$ is nonzero only in vicinity of the Fermi level~\cite{Leyland07,Griesbeck09}. Then, the spin polarization is determined by the effective field and isotropization time at the Fermi level. Equation~(\ref{kinetic}) yields 
\begin{equation}\label{eq:main}
{\bf s_k} = \frac{ \boldsymbol{\zeta}_{\bf k} +  \tau \Ok \times \boldsymbol{\zeta}_{\bf k} + \tau^2 \Ok (\Ok \cdot \boldsymbol{\zeta}_{\bf k})}{1 + \Ok^2 \tau^2} \:,
\end{equation}
where $\boldsymbol{\zeta}_{\bf k} = \av{\bf s_k} + {\bf g_k}\tau$ and $\Ok = \boldsymbol{\Omega}_L + \boldsymbol{\Omega}_{{\bf k}}$.
The closed equation for $\av{\bf s_k}$ can be found by averaging both sides of Eq.~(\ref{eq:main}) over the wave vector directions~\cite{Tarasenko06}. For the case where ${\bf g_k}$ is independent of the direction of $\bf k$, ${\bf g_k} = \bf g$, such a procedure gives
\begin{equation}\label{eq:main_av}
[(\av{\bf s_k} + {\bf g}\tau ) \times \tilde{\bf \Omega}]_{\alpha} = g_{\alpha} - \sum_{\beta} \tilde{\Gamma}_{\alpha \beta} (\av{\bf s_k}_\beta + g_\beta \tau ) \:,
\end{equation}
where
\begin{eqnarray}\label{eq:eff}
\tilde{\bf \Omega} = \av{\frac{\Ok}{1 + \Ok^2 \tau^2}} \:, \;\; \tilde{\Gamma}_{\alpha \beta} = \av{\frac{\Ok^2\delta_{\alpha \beta} - \Omega_\alpha \Omega_\beta}{1 + \Ok^2 \tau^2}}\tau \:.
\end{eqnarray}

In $(001)$-grown quantum wells, the effective magnetic field is linear in the wave vector and consists of the Rashba and Dresselhaus contributions~\cite{Averkiev08}. The corresponding frequency $\Ok_{\bf k}$ can be presented in the form 
\begin{equation}\label{eq:splitting}
\Ok_{\bf k} =  \left[ (\Omega_D+\Omega_R)k_y/k_F,\: (\Omega_D-\Omega_R)k_x/k_F,\: 0 \right]\:, 
\end{equation}
where $\Omega_D$ and $\Omega_R$ are the precession frequencies in the Dresselhaus and Rashba effective fields at the Fermi level, respectively, $k_F$ is the Fermi wave vector, $x\parallel [1\bar{1} 0]$ and $y\parallel [110]$ are the in-plane axes. Note, that the $g$-factor tensor is diagonal in this coordinate frame. Solution of Eq.~(\ref{eq:main_av}) for arbitrary parameters $\Omega_D$ and $\Omega_R$ is complicated and can be found numerically. However, it has the simple form provided the frequency magnitude $|\Ok_{\bf k}|$ is independent of the direction of $\bf k$ which occurs if either $\Omega_D$ or $\Omega_R$ is zero. In particular, in the case of Rashba splitting
and the spin generation along the QW normal, 
the out-of-plane and in-plane components of the total electron spin ${\bf S} = \sum_{\bf k} \bf s_k$ have the form
\begin{widetext}
\begin{equation}\label{S_z}
S_z = \frac{\Omega_R^2 \tau \, G_z}{(\Omega_R^2-\Omega_L^2)^2 \, \tau^2 + \Omega_L^2 \left[1+\sqrt{1+(\Omega_R+\Omega_L)^2\tau^2}\sqrt{1+(\Omega_R-\Omega_L)^2\tau^2}\right]} \:,
\end{equation}
\vspace{-2mm}
\begin{equation}\label{S_parallel}
{\bf S}_{\parallel} = \frac{4 \boldsymbol{\Omega}_L \times {\bf G}} {4\Omega_L^2+(\Omega_R^2-\Omega_L^2)^2\tau^2 + (\Omega_R^2-\Omega_L^2)\left[\sqrt{1+(\Omega_R+\Omega_L)^2\tau^2}\sqrt{1+(\Omega_R-\Omega_L)^2\tau^2}-1\right]} \:,
\end{equation}
\end{widetext}
where ${\bf G} = \sum_{\bf k} \bf g$ is the spin generation rate. Similar equations for the Dresselhaus splitting are obtained from Eqs.~(\ref{S_z}) and~(\ref{S_parallel}) by replacing $\Omega_R$ with $\Omega_D$. 

Shown in Figs.~1(a) and~(b) are the magnetic field dependences of the spin components $S_z$ and $S_x$ plotted
for ${\bf B} \parallel y$ and different products $\Omega_R\tau$. At $\Omega_R\tau \ll 1$, the field dependences of the spin components are described by classical Hanle curves 
\begin{equation}\label{S_collision}
S_z = \frac{T_z G_z }{1+\Omega_L^2 T_z T_{\parallel}} \:, \;\;\; {\bf S}_{\parallel} = \frac{T_z T_{\parallel} \, \boldsymbol{\Omega}_L \times {\bf G}} {1+\Omega_L^2 T_z T_{\parallel}} \:,
\end{equation}
which also follow from Eqs.~(\ref{S_z}) and~(\ref{S_parallel}). Here, $T_z = 1/(\Omega_R^2 \tau)$ and $T_{\parallel} = 2/(\Omega_R^2 \tau)$ are the out-of-plane and in-plane spin relaxation times in the collision-dominated regime~\cite{DK86}. The dependences $S_z(B)$ and $S_x(B)$ given by Eqs.~(\ref{S_collision}) are shown by dashed curves. In QWs with high electron mobility and strong spin-orbit coupling, where the parameter $\Omega_R\tau$ is large, the magnetic field dependence of the electron spin drastically changes. 
Instead of a monotonic decrease with the field, $S_z(B)$ rises first with the field, reach maximum, and decreases as $1/B^4$ at high fields, see Eq.~(\ref{S_z}) and Fig.~1(a). At $\Omega_R\tau \gg 1$, the Hanle curve $S_z(B)$ reaches maximum in the magnetic field where $\Omega_L=\Omega_R$. This can be used in experiment for the direct measurement of the effective magnetic field. The electron spin $S_z$ reaches the value $G_z/(2\Omega_R)$ at $\Omega_L=\Omega_R$ that is independent of the scattering time and much larger then $S_z(0)=G_z/(\Omega_R^2 \tau)$. We also note that the field dependence of the in-plane spin component changes qualitatively with the increase of $\Omega_R\tau$ as well [Fig.~1(b)].
In high-mobility QWs, $S_x(B)$ exhibits the sharp rise at $\Omega_L \approx \Omega_R$ and exceeds $S_z(0)$, which does not occur in the collision-dominated regime. 
\begin{figure}[b]
\includegraphics[width=0.9\linewidth]{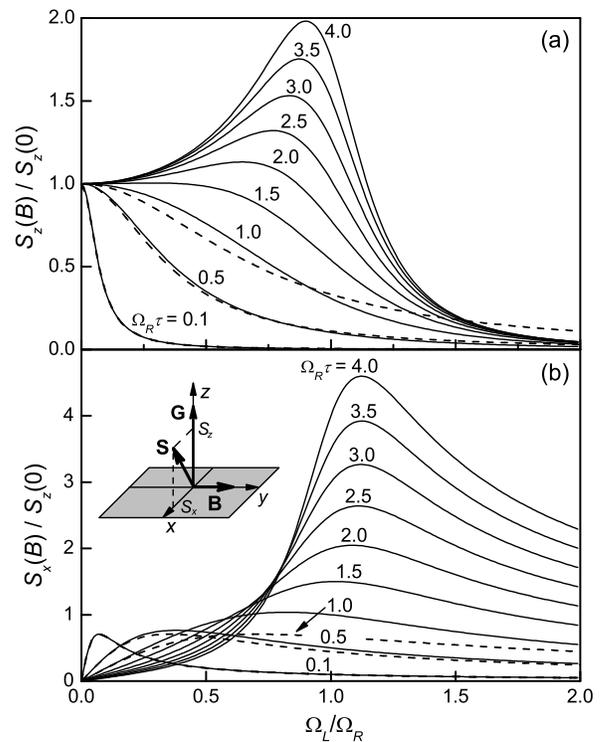}
\caption{Dependences of $S_z$ and $S_x$ on the magnetic field ${\bf B} \parallel y$ plotted after Eqs.~(\ref{S_z}) and~(\ref{S_parallel}) (solid curves) and classical Eqs.~(\ref{S_collision}) (dashed curves) for different $\Omega_R \tau$. Inset in panel (b) illustrates the geometry of experiment.}
\end{figure}

Equations~(\ref{S_z}) and~(\ref{S_parallel}) allow us to analyze the Hanle curves 
at small fields. To second order in $\boldsymbol{\Omega}_L$, the field dependences of the spin components have the form
\begin{equation}\label{S_smallB}
S_z \approx \left[1- \Omega_L^2 T_z T_{\parallel} \left(1- \frac{\Omega_R^2\tau^2}{2} \right)\right] T_z G_z \:, 
\end{equation}
\vspace{-5mm}
\[
{\bf S}_{\parallel} \approx T_z T_{\parallel} \,\boldsymbol{\Omega}_L \times {\bf G} \:.
\]
One can see that $d S_z /d B =0$ at zero magnetic field while the sign of $d^2 S_z /d B^2$ is determined by $\Omega_R \tau$. The component $S_z$ decreases or increases with the field induction if $\Omega_R \tau < \sqrt{2}$ or $>\sqrt{2}$, respectively. Such a transition from classical to anomalous Hanle effect is clearly seen in Fig.~1(a). However, the deviation of $S_z(B)$ from the Lorenz function (\ref{S_collision}) is already considerable for $\Omega_R \tau =1$.  

The increase of $S_z$ in the in-plane magnetic field and its maximum at $\Omega_L = \Omega_R$ in high-mobility structures can be understood by analyzing the dynamics of spin polarized electrons in the external and effective magnetic fields. Figure~2 illustrates the distributions of the frequency $\Ok = \boldsymbol{\Omega}_L + \boldsymbol{\Omega}_{R}$, which determines the spin precession of individual electrons, in ${\bf k}$-space. Panel~(a) corresponds to $\boldsymbol{\Omega}_L=0$, panel~(b) corresponds to $\boldsymbol{\Omega}_L \parallel y$ and $|\boldsymbol{\Omega}_L|=|\boldsymbol{\Omega}_R|$. In the absence of external magnetic field [Fig.~2(a)], the electron spins initially oriented along the QW normal rapidly precess in the effective field with the frequency $\Omega_R$~\cite{Leyland07,Griesbeck09,Gridnev01,Culcer07,Glazov07}. Such a precession results in a small time-average spin $S_z(0) = G_z/(\Omega_R^2 \tau)$ that is measured in the regime of continuous-wave pumping, where the generation rate ${\bf G}$ is constant on the spin lifetime scale. The external in-plane magnetic field ${\bf B}$ changes the magnitude and direction of $\Ok$ thereby affecting the spin dynamics. In particular, the external field with the Larmor frequency $|\boldsymbol{\Omega}_L| = |\boldsymbol{\Omega}_R|$ compensates the effective field in a certain point at the Fermi circle. This is the point ${\bf k}_0=(k_F,0)$ in Fig.~2(b). In the vicinity of ${\bf k}_0$, $\Ok$ vanishes and the electron spins do not precess conserving their orientation. Such a suppression of spin dephasing results in the increase of $S_z$ at $|\boldsymbol{\Omega}_L| \approx |\boldsymbol{\Omega}_R|$ in high-mobility QW structures.
\begin{figure}[t]
\includegraphics[width=0.85\linewidth]{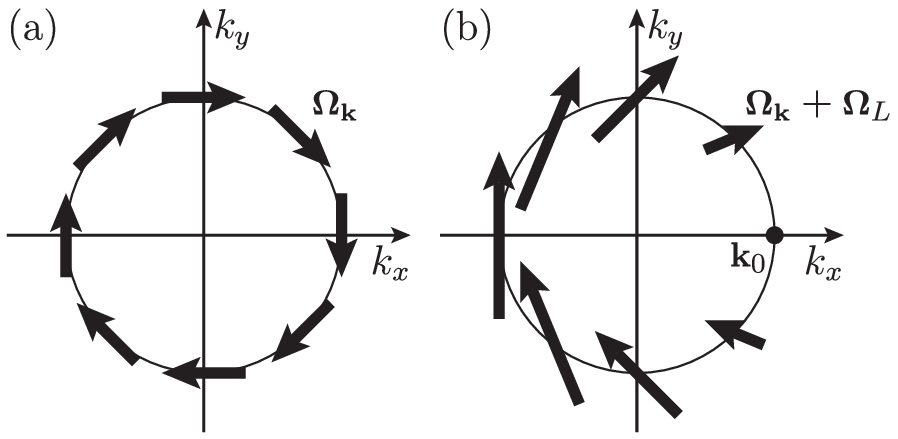}
\caption{Distributions of the frequency $\Ok = \boldsymbol{\Omega}_L + \boldsymbol{\Omega}_{R}$ 
in ${\bf k}$-space for two particular cases: (a) $\boldsymbol{\Omega}_L=0$ and (b) $\boldsymbol{\Omega}_L \parallel x$ and $|\boldsymbol{\Omega}_L|=|\boldsymbol{\Omega}_R|$.}
\end{figure}

Generally, the effective magnetic field in asymmetric (001)-grown QWs consists of both Rashba and Dresselhaus contributions. It leads to an in-plane anisotropy of spin properties such as spin relaxation rate and effective $g$-factor~\cite{Averkiev06,Stich07,Meier08,Larionov08,Eldridge11}. Our analysis shows that the spin dynamics in asymmetrical quantum wells is similar to that considered above. Depending on the strength of spin-orbit splitting, $\boldsymbol{\Omega}_{\bf k} \tau$, the Hanle effect in the in-plane magnetic field demonstrates classical or anomalous behavior. However, the Hanle curve depends on the orientation of the field ${\bf B}$ if both Rashba and Dresselhaus terms are present. Moreover, it can happen that $S_z$ monotonously decreases with the field for some orientations of ${\bf B}$ while it exhibits nonmonotonic behavior for other ${\bf B}$ orientations. An example of such strongly anisotropic dependence of $S_z$ on the in-plane field is shown in Fig.~3. In this particular case, $S_z({\bf B})$ is a monotonic function for ${\bf B}\parallel y$ and nonmonotonic function for ${\bf B}\parallel x$.
\begin{figure}[b]
\includegraphics[width=0.9\linewidth]{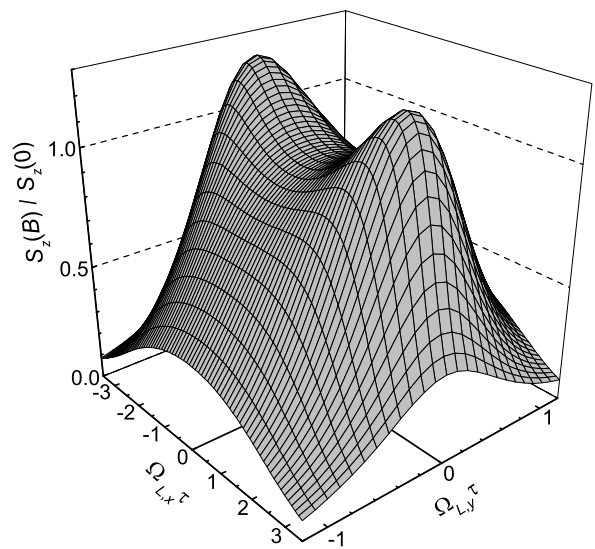}
\caption{Dependence of $S_z$ on the in-plane magnetic field of different orientation for $\Omega_D \tau = 2$ and $\Omega_R \tau = 1$.}
\end{figure}

In high-mobility QW structures, the dependence $S_z$ on the in-plane field has a sharp maximum at $\Ok_L \approx \Ok_L^*$ where the frequency $\Ok = \Ok_L+\Ok_R+\Ok_D$ vanishes for a certain wave vector at the Fermi circle. In the limit $\Omega_R\tau, \Omega_D\tau \gg 1$, the  frequency $\Ok_L^*$ is given by  
\begin{equation}
{ 1 \over \Omega_L^{*}} \approx \sqrt{ \frac{\cos^2 \phi}{(\Omega_R+\Omega_D)^2} + \frac{\sin^2 \phi}{(\Omega_R-\Omega_D)^2} } \:,
\end{equation}
where $\phi = \arctan (\Omega_{L,y}/\Omega_{L,x})$ is the angle between $\Ok_L$ and the $x$ axis. The out-of-plane spin component $S_z$ in this magnetic field reaches the value
\begin{equation}
S_z (\Omega_L^*) \approx  \frac{G_z}{2} \sqrt{ \frac{(\Omega_R+\Omega_D)^2\sin^2 \hspace{-1mm} \phi + (\Omega_R-\Omega_D)^2\cos^2 \hspace{-1mm} \phi}{(\Omega_R+\Omega_D)^4\sin^2 \hspace{-1mm} \phi + (\Omega_R-\Omega_D)^4\cos^2 \hspace{-1mm} \phi} } 
\end{equation}
independent of the relaxation time $\tau$. The study of the angle dependences of $\Omega_L^*$ and $S_z (\Omega_L^*)$ in experiment can help in determining the coefficients of spin-orbit splitting of electron subbands in QWs.   

To summarize, we have studied the spin dynamics of optically oriented electrons in quantum wells in the in-plane magnetic field. It has been shown that the interplay of spin-orbit and external magnetic fields leads to an anomalous Hanle effect in high-mobility structures. We have derived an analytical expression for the Hanle curves in structures with arbitrary strength of spin-orbit coupling, which can be used for fitting experimental data and determining the spin-orbit magnetic field.

\paragraph*{Acknowledgments.} This work was supported by the RFBR, programs of the RAS, and the Foundation ``Dynasty''-ICFPM.

\end{document}